\documentclass[journal]{IEEEtran}

\usepackage{graphicx} %package to manage images
\graphicspath{ {./Images/} }

\usepackage{amsmath}
\usepackage{amsthm}
\usepackage{amssymb}

\usepackage[english]{babel}

\ifCLASSINFOpdf
\else
\fi

\begin{document}

\title{Simple HF antenna efficiency comparisons using the WSPR system}

\author{Jens~Zander,~\IEEEmembership{Life-Member,~IEEE,}
\thanks{J. Zander is with the Department
of Electrical Engineering and Computer Science, KTH-Royal Institute of Technology, S-10044 Stockholm, Sweden}}

\maketitle

\begin{abstract}

Determining the efficiency of an HF-antenna by measurements requires is a complex procedure involving expensive equipment, calibrated instruments for field strengths. In this paper we evaluate a simple, inexpensive method to determine the relative efficiency of an antenna relative a reference antenna. The method uses the \emph{Weak Signal Propagation Reporter(WSPR)} network of receivers that are located all over the world. These receivers report the estimated signal-to-noise ratio of received beacon signals to the WSPR.net database where the data can be retrieved (almost) in real time. In the paper we analyze the method, estimate its accuracy and discuss advantages and limitations. Some preliminary measurement results are presented.
\end{abstract}


\section{Introduction}

Experimentally verifying  far field performance of an antenna for the HF frequency range (2-30 MHz) such as antenna gain and and antenna efficiency is a complex procedure that involves expensive equipment and accurately calibrated instruments. The antenna radiation or gain pattern of an antenna \cite{Balanis}
\begin{equation} \label{antgain}
G(\phi,\theta)=\eta D((\phi,\theta)
\end{equation}
is composed of a part \(D\), the \emph{directivity or directivity pattern} that is dependent on the direction of radiation, and a part \(\eta\)  that is independent of the direction. The latter is normally denoted the \emph{efficiency}.  As the antenna directivity pattern sums up to unity when integrating over all direction, is fairly easy to measure with simple,  non-calibrated instruments since only the relative field strengths needs to be measured. When it comes the efficiency,  absolute power or field strength instruments have to be measured requiring calibrated instruments.  For evaluating microwave antennas, drone based measurement systems that fly around the antenna measuring the field strength, are now commercially available, e.g. \cite{DRONE}. Such systems are, however, very expensive and complex.  In addition,  HF antennas, in particular for the lower HF frequency bands,  are large constructions and physically measuring the field strength in the far field of an antenna will be quite impractical, in particular in the elevation domain.  Using skywave signals from distant transmitters has also the drawback that the propagation loss varies and is not accurately known.

The \emph{Weak Signal Propagation Reporter(WSPR)} \cite{Taylor} designed by Joe Taylor(K1JT) was introduced in 2010 and the protocol has gained significant interest in the Amateur radio community since then. Hundreds of transmitters and receivers are using the protocol to constantly monitor HF-propagation conditions in the HF amateur bands. All signal reports are sent to the WSPR.net repository where they can be easily accessed and later archived for future research purposes.
Several methods for antenna measurements relying on the WSPR network have been proposed. In its simplest form, a WSPR transmitter, is connected to the antenna of interest. The measurements of the signal quality received at certain distant receivers are then used to assess the performance of the antenna. The DXplorer application by SOTABEAMS \cite{DXPLORER} is an example of a proprietary application that will process the WSPR data from the WSPR.net data base and present various performance metrics that allow for rough comparisons of antenna performance. A drawback is that what is measured is the combined effect of propagation loss and antenna performance.  Another method, more specifically targeting the antenna performance is described in \cite{160M} where the performance of two antennas in the 160m amateur band is compared. Here, the performance of the antenna to be measured is compared to a known reference antenna. A remote WSPR transmitter is received with two parallel receivers, one for each antenna. If the antennas are located relatively close to each other, the propagation loss from the antenna to both receiving antennas is the same, and we can make a direct comparison between the performance of the two antennas. The method requires that the two receivers are calibrated for identical performance. We could repeat the procedure, receiving the signal from many different stations to get a more comprehensive coverage of the antenna gain pattern.

In this paper we propose a simple, inexpensive method to determine the antenna performance relative to a reference antenna. The method can be seen as a combination of the methods in \cite{DXPLORER} and \cite{160M}. Like in \cite{DXPLORER} we use a simple, inexpensive, WSPR transmitter connected to the antenna to be measured. However, we also use a second transmitter connected to a reference antenna. We collect the reported signal qualities from the WSPR.net database for both antennas. If the signal quality from both antennas is reported by the same station in the same time interval, we can assume that the propagation loss is the same to both antennas and a direct comparison of the antenna performance is possible. In the paper, this method is analyzed and its accuracy is estimated. We will further discuss the advantages and disadvantages of this method compared to the two-receiver technique in \cite{160M}.

\section{The WSPR system}
The WSPR protocol was originally designed by Joe Taylor \cite{Taylor}. WSPR stations transmit short messages of only 50 bits containing 
\begin{itemize}
    \item the call sign of the transmitting (amateur radio) station, 
    \item The geographical location of the transmitting stations, coded as a 6-position Maidenhead locator \cite{Maidenhead}, and
    \item the power level of the transmitter in dBm.
\end{itemize}

The message in the standard (WSPR-2) protocol is heavily coded (4-FSK, 1.46 symbols/s) and transmitted in a 2 minute frame, occupying only 6 Hz bandwidth allowing reception down to an SNR of -28dB (using a 2500Hz reference bandwidth) \footnote{There is also a very weak signal mode, WSPR-15, that uses 15 minute transmission intervals and works down to -37dB SNR}. The duration of each transmission is 110.6 seconds and starts one second into an even Universal Time Coordinated (UTC) minute. The remainder of the 2 minute frame duration can be used for decoding of the messages. There are several publicly available software decoders. The most commonly used is the WSJT-X software package \cite{WSJTX} that also contains support for several other weak-signal digital transmission modes.

The receivers continuously report the received signal reports to the WSPR.net site \cite{WSPRNet} from were the information can be retrieved. The database contains the following key items for each signal report
\begin{itemize}
    \item Timestamp
    \item Callsign of transmitter
    \item Frequency
    \item SNR estimate in dB (rounded to integer)
    \item Transmitter power in dBm  (rounded to integer)
    \item Transmitter location (6 character Maidenhead geolocator)
    \item Callsign of reporter
    \item Reporter location (6 character Maidenhead geolocator)
\end{itemize}
In addition the estimated frequency drift (Hz/s) and the transmission mode (typically 2 for WSPR-2) are stored. Also the distance and azimuth are calculated and stored in the database

\begin{figure}[t]
\centering
\includegraphics [width=8.5cm]{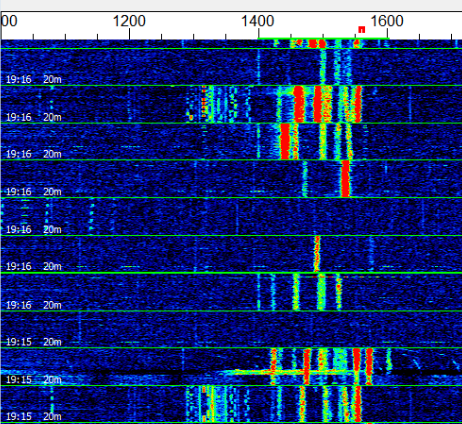}
\caption{WSPR spectrum display from \cite{WSJTX}. WSPR signals are found in the baseband equivalent interval 1400 - 1600Hz. Horizontal lines indicate 2 minute intervals, the colors indicate the received power}
\label{fig:Waterfall}
\end{figure}

The signals are transmitted in specific 200Hz wide frequency segments in each of amateur radio HF bands from 1.8MHz (160m) to 28MHz (10m).  Here, the WSPR transmitters chose a 6Hz "slot" for their transmission, which is illustrated in figure \ref{fig:Waterfall}.  Usually, there are many WSPR-transmission in progress simultaneously,  so there is a risk of "collisions" occurring at the receiver. Under favorable conditions, at most 30 (approx 200/6) simultaneous signals could be received simultaneously without serious signal degradation. If two or more signals are received that are too close in frequency, the decoder may in some cases not be able to decode one or more signals. In these cases, either no report at all or an unreliable SNR report may result. 

Simple receivers using the WSJT-X software are able to monitor one frequency band at the time. Many of them "hop" between bands between different transmission intervals to get a comprehensive view of current propagation conditions. Some of the receiving stations may also transmit themselves in certain transmission intervals. This means that there may be a significant time lapse between two consecutive reports from such a receiver.  On the other hand, dedicated and direct-sampling  digital receivers are capable of monitoring all frequency bands simultaneously. Such receivers may produce a continuous stream of signal reports that is only interrupted by interference or severe fading events.

\begin{figure}[ht]
\centering
\includegraphics [width=8.5cm]{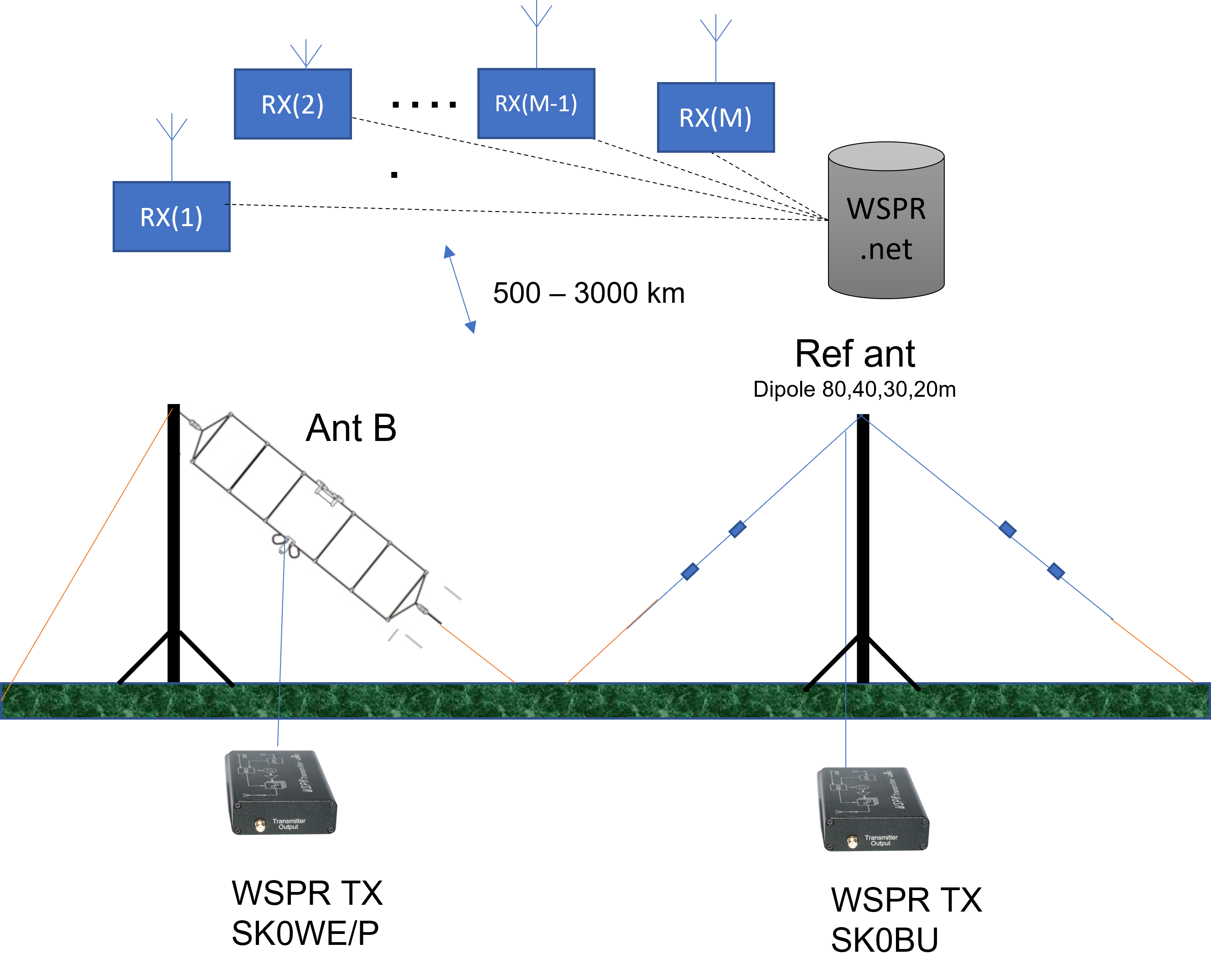}
\caption{Measurement setup (Experiment B: T2FD vs Dipole @ 7MHz)}
\label{fig:Setup}
\end{figure}

\begin{figure}[ht]
\centering
\includegraphics [width=7cm]{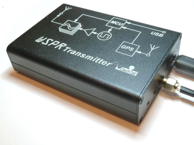}
\caption{Zachtek 200mW WSPR transmitter \cite{Zachtek1}}
\label{fig:Zachtek}
\end{figure}

\section{Measurement setup}
Figure \ref{fig:Setup} describes the measurement setup. One WSPR-transmitter is connected to each of the antennas: the antenna to be measured (Ant B) and the reference antenna respectively. The WSPR transmitters are configured to transmit in the same frequency band during the same 2 minute time slot. Each transmitter uses a separate callsign to allow for the separation of signals at the receiver. There are several, low cost,  standalone WSPR-transmitters on the market. The one used in the experiments was a Zachtek transmitter \cite{Zachtek1} shown in fig \ref{fig:Zachtek}. This device is USB-powered and also includes a GPS-receiver for determination of the transmitter location as well as for providing an accurate time-reference for the WSPR protocol. The output power is fixed to 200mW (+23dBm)
Focusing only on one particular frequency band, signals in timeslot \(k\) are received by a random set of WSPR-receivers \(RX_1 ...RX_M^{\ast} \) located "all over the world" that report the estimated signal-to-noise ratio (SNR) to from each received transmission to the WSPR.net database. From the database, we extract signal reports from those receivers \(RX_i\) that have been been receiving signals from both antennas in time slot \(k\) for further processing. Note that there is no need for the receivers to be calibrated. The noise environments at the receiver sites may be very different, resulting in different SNR readings for the same received signal power. However, as the signals from both antennas are received in same receiver, the difference between the SNRs is independent of the propagation loss and noise level at the receiver.  The SNR differences can be averaged for a specific receiver station to estimate the antenna gain pattern in relation to the reference antenna for that specific azimuth and elevation, or over all receiver stations to estimate the efficiency of the antenna. The properties of these estimates are described in the following section.

\section{Estimation of antenna parameters}
\subsection{Modelling assumptions}
Our aim is to estimate the antenna gain pattern and the antenna efficiency based on the reported signal-to-noise ratios from receiver \(i\) in time slot \(k\). For the antenna under measurement we denote this SNR \(\Gamma_{i,k}\) and \(\hat{\Gamma}_{i,k}\). The SNR measured at receiver \(i\) from the measurement antenna in timeslot \(k\) can be written as

\begin{equation} \label{RXSNR}
    \Gamma_{i,k} = \frac{P_t G(\phi_i,\theta_i)G_{rx,i}}{L_{i,k}(N_i+I_{i,k})}
\end{equation} 
where \(\phi_i\) and \(\theta_i\) denote the azimuth and elevation of the wave leaving the antenna on its way to receiver \(i\), \(L_{i,k}\) denotes the pathloss to receiver to receiver \(i\), \(N_i\) is the background noise at receiver \(i\) and \(I_{i,k}\) is the interference (from other WSPR users) disturbing the reception at at receiver \(i\) in timeslot \(k\). The transmit power \(P_t\) is assumed to be the same in both antennas as identical WSPR-transmitters are used.  The SNR from the reference antenna has the same expresion as in eq \ref{RXSNR}, where we replace \(G(\phi_i,\theta_i)\) with \(\hat{G}(\phi_i,\theta_i)\). Further, we need to replace \(I_{i,k}\) with \(\hat{I}_{i,k}\) we use two different frequencies for our signals in the 200kHz segment so the the narrowband interference from other WSPR-transmitters will affect the two received signals independently. On other hand we may assume that the pathlosses \(L_{i,k} = \hat{L}_{i,k} \) since the 200kHz frequency band is very narrow compared to the typical correlation bandwidth of HF channels, resulting in "flat" fading over the entire frequency segment. The pathloss will however be time-varying, i.e. will depend on \(k\). Also the receiver antenna gain \(G_{rx,i}\) is assumed to be constant.

\subsection{Antenna gain pattern estimation}

Now the ratio between the two SNRs can be written as (using equation \ref{antgain})
\begin{equation} \label{linrat}
    \frac{\Gamma_{i,k}}{\hat{\Gamma}_{i,k}} = 
    \frac{ G(\phi_i,\theta_i)}{\hat{G}(\phi_i,\theta_i)} \times \frac{1+\frac{\hat{I}_{i,k}}{N_i}}{1+\frac{I_{i,k}}{N_i}}
\end{equation}

Expressing the SNR ratio in dB we get

\begin{equation} \label{lograt}
\begin{split}
   & \Delta S_{i,k} = S_{i,k}-\hat{S}_{i,k} = \\
   & = (G(\phi_i,\theta_i)_{dB}-\hat{G}(\phi_i,\theta_i)_{dB}) \\
   & + \epsilon_{i,k}
    \end{split}
\end{equation}
where \(S_{i,k}\) and \(\hat{S}_{i,k}\) are the reported SNR:s for the measured antenna and the reference antenna in dB. The error term can be written as
\[\epsilon_{i,k}=10lg(\frac{1+\frac{\hat{I}_{i,k}}{N_i}}{1+\frac{I_{i,k}}{N_i}}) \]

As the background noise power \(N_i\) is the same for both received signals at recevier \(i\), we will assume that both \(X=\frac{I_{i,k}}{N_i}\) and \(\hat{X}=\frac{\hat{I}_{i,k}}{N_i}\) are independent and identically distributed. Computing the expectation 

\begin{equation} \label{Experror}
\begin{split}
& E[\epsilon_{i,k}]=E[10lg(\frac{1+X}{1+\hat{X}})] \\
& = E[10lg(1+X)] - E[10lg(1+\hat{X})]) = 0   
\end{split}
\end{equation}

We can conclude that \(\Delta S_{i,k}\) is an unbiased estimator of \(G(\phi_i,\theta_i)_{dB}-\hat{G}(\phi_i,\theta_i)_{dB})\). This also holds for the average estimator, where we average over \(K\)  SNR samples from receiver \(i\)
\[\Delta S_i =\sum_{k=1}^{K} \Delta S_{i,k} \]
that has a variance of that decreases as \(\sigma^2/K\) where \(\sigma^2\) is the variance of \(\epsilon_{i,k}\)

Although it is clear that we can achieve arbitrary accuracy (low variance)  by taking "enough" samples, it would be interesting to know, how many samples are needed to achieve a certain variance.  Estimating \(\sigma^2\) is more complex as the error will depend on the received power (weak signals are more easily disturbed than strong ones) and the fact that strong interference will result in a decoding failure which will, in turn, not generate a SNR report at all. Also the quantization effects (to whole dB) and the SNR estimation algorithm used in the receiver software need to be considered\footnote{Most receivers reconstruct the transmitted signal from the successfully decoded data and compare this with the received signal, e.g. by computing the mean-square error which is mapped to an SNR by table lookup} . More model assumptions and a detailed study of the decoder behavior is required. As an alternative the variance can be estimated from the measurements themselves as discussed in section \ref{experiments}.

A difficulty when we are interested in estimating the whole gain pattern,  is of course that we will get measurements only for a given set of \((\phi_i,\theta_i)\) corresponding to the set of locations of the active receivers. Specifically, we will have difficulty in estimating the pattern close to nulls as we will get no reports from those directions.

\subsection{Antenna efficiency estimation}

We are now interested in determining if 

\begin{equation} \label{estimator}
   \Delta S_k =\frac{1}{M_k^{\ast}}\sum_{i=1}^{M_k^{\ast}} \Delta S_i
\end{equation}

The summation is here taken over the \(M_k^{\ast}\) receivers that are providing reports for\emph{both} antennas in time slot \(k\). As we noted in the previous section, this number will be a random number that varies from time slot to time slot. We will now investigate if expression \ref{estimator}  a usable estimator of the difference in antenna efficiencies 
\[\Delta \eta =\eta_{dB} -\hat{\eta}_{dB}\]

Returning to eq. \ref{antgain} and eq. \ref{linrat} we can write
\begin{equation} \label{logeff}
\begin{split}
   & \Delta S_k = S_k-\hat{S}_k = \Delta \eta\\
   & + \sum_{i=1}^{M_k^{\ast}} (D(\phi_i,\theta_i)_{dB}-\hat{D}(\phi_i,\theta_i)_{dB}) \\
   & + \sum_{i=1}^{M_k^{\ast}} \epsilon_{i,k}
    \end{split}
\end{equation}
Taking the expectation yields

\begin{equation} \label{expeff}
E[\Delta S_k] = \Delta \eta + E[\sum_{i=1}^{M} (D(\phi_i,\theta_i)_{dB} -\hat{D}(\phi_i,\theta_i)_{dB})]
\end{equation}
as \(E[\epsilon_{i,k}]=0\) and assuming that all \(M\) active receivers are equally likely to provide reports. 

As we can see from eq. \ref{expeff} is that the estimator \(\Delta S\) will have a bias
\begin{equation} \label{bias}
   B= E[\sum_{i=1}^{M} [D(\phi_i,\theta_i)_{dB}] -E[\sum_{i=1}^{M}\hat{D}(\phi_i,\theta_i)_{dB})]
\end{equation}
where the expectation is taken over the random receiver directions. Unfortunately there is no closed-form expression of this expectation as we are summing the directivities in dB (and not in linear scale). One can however see,  that if the antennas have similar directivity, e.g. if they are approximately omnidirectional, \(B \approx 0 \) and thus will \(\Delta S\)  be an unbiased estimator of \(\Delta \eta\). Whenever there is a significant difference between the directivity patterns, the estimator may become biased. If the directivity pattern of the reference antenna is known, we can estimate the bias in eq. \ref{expeff} by using the gain pattern estimates in the previous subsection, using the fact that the directivity should add up to unity for a uniform spatial distribution of receivers.  

A more significant problem arises if the spatial distribution of receivers is not uniform as this may add bias to the estimate. For example, if most of the receivers are found in a certain range of directions, differences in directivity between the measured antenna and the reference antenna will create such bias. In this case one should consider repeated measurements, changing the orientation of the measured antenna.

\section{Preliminary Field experiments} \label{experiments}

The method was tested on different frequency bands with different antennas. Here, we will presents results from experiments with two types of antennas:

\textbf{Experiment A:}  Comparison of commercial vertical antennas. 14 MHz daylight, spring conditions. Location: Gotland Island (JO97)
\begin{itemize}
        \item \textbf{Antenna A}:Difona HF-P1 portable vertical - a short (2.5m) center loaded resonant vertical with 4 radials \cite{Kuhl} 
        \item \textbf{Reference A}: Hy-Gain AV620 Patriot - resonant 5/8 \(\lambda\) vertical
 \end{itemize}
 
\textbf{Experiment B:} Evaluation of Terminated Tilted Folded Dipole (T2FD).  7 MHz daylight, spring conditions. Location: Stockholm (JO89)
\begin{itemize}
        \item \textbf{Antenna B}: Diamond WD-330 T2FD Aperiodic broadband dipole 25m long, high point 15m 
        \item \textbf{Reference B}: "Inverted V" dipole 1/2 \(\lambda\) high point 10m
\end{itemize}

\begin{figure}[t]
\centering
\includegraphics [width=8.5cm]{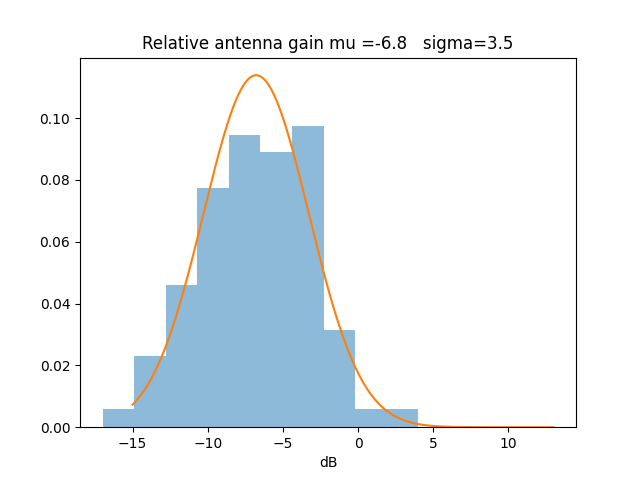}
\caption{Histogram of measured  \(\Delta S_k\) (dB). Short center-loaded vertical (Difona) at 14 MHz - reference: Hy-Gain AV620  5/8 \(\lambda\) vertical}
\label{fig:Difona14}
\end{figure}

\begin{figure}[ht]
\centering
\includegraphics [width=8.5cm]{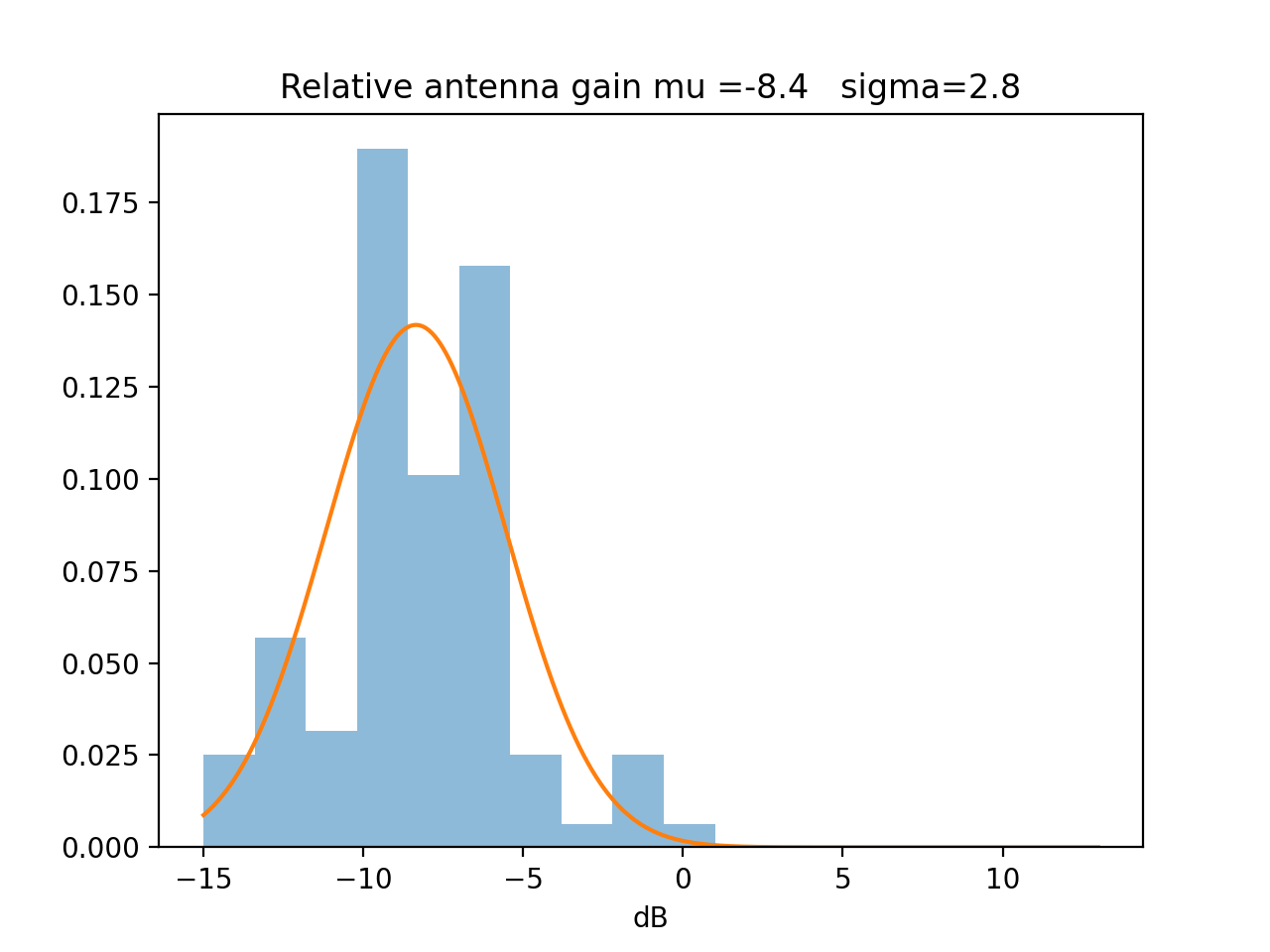}
\caption{Histogram of measured  \(\Delta S_k\) (dB). 25m T2FD - reference 1/2 \(\lambda\)dipole}
\label{fig:T2FD7}
\end{figure}

In each experiment, about 1000 signal reports were collected during roughly 1 hour. The signal reports were retrieved from the WSPR.net web-interface after the measurement period, The SNR reports were then processed using a simple Python script, filtering out the valid signal reports, i.e where signals from both antennas were received at same receiver. The result was 150 - 200 valid signal reports from 15 - 35 different receiving stations.  

The results are presented in fig \ref{fig:Difona14} and fig \ref{fig:T2FD7} in the form histogram plots for the measured \(\Delta S_k\). The number of reporting stations was lower a 7 MHz due to propagation conditions whereas the congestion is likely to be more noticeable at 14 MHz. The figures also show the estimated (average) difference \(mu=\Delta \eta)\) and the estimated sample standard deviation \(\sigma\). Noticeable is that \(\sigma\) in these (and other experiments) is estimated close to 3 dB. The standard deviation for the mean estimate is \(\frac{\sigma}{\sqrt{\hat{K}}}\) where \(\hat{K}\) is the number useful samples. In our case with \(\hat{K} \approx 100\) the standard deviation of the average estimator would be less than 0.5dB.  The efficiency results from experiment B have been compared with NEC2(EZNEC) simulations\cite{Malhotra} and are found to be roughly consistent.

The number of reporting stations was lower at 7 MHz due to propagation conditions whereas the congestion is likely to be more noticeable at 14 MHz. The bulk of the signal reports on 14 MHz came from stations with at a distance of 1500-2000 km and in the 7 MHz the typical distance is 500-1000 km. WSPR receivers are more frequent in Western Europe, which means that most of the SNR reports in Sweden are likely to come from directions in the south-west quadrant. In the experiments, this is not likely to be a significant problem as fairly omnidirectional antennas were compared. 

\section{Discussion}
In this paper we have discussed a simple, low cost, "crowdsourcing" technique for comparing the performance of HF antennas using the widespread and global network of WSPR receivers. In our method, we transmit WSPR signal over the antennas to be measured, instead of connecting them to receivers as in \cite{160M}. This has the advantage that their are many more receivers active that WSPR transmitters. Further, the method does not require any receiver calibration (as in \cite{160M}) and by comparing with signals from a reference antenna we compensate for variations in receiver noise floor and propagation fluctuations.  By using long measurement periods (hours) we can also strongly reduce the effects of temporary congestion at the WSPR receivers \footnote{A potential problem with long measurements of the antenna gain is that the propagation mode may change during the measurements, which may effect the angle of incidence \(\theta_i\)}.  Our analysis and measurements show that the antenna gain relative to the reference antenna can be estimated within an accuracy of less than a dB within a few hours measurements.The drawback is that we are limited to estimating the gain pattern in those directions were there are receivers. Estimating the antenna efficiency relative to the reference antenna can be done with by averaging the relative SNR differences over all receivers. This estimator, however, may have a bias, if the gain patterns of the measured antenna compared to the gain pattern of the reference antenna exhibits large differences, in particular in those directions where many receivers are found.  For approximately omnidirectional antennas this bias may be manageable. 

The main error source and the key drawback of the method is the uneven distribution of receivers. In our case most of the receivers are located in Western Europe, which concentrates our measurements to southern to western azimuths. The only way could be counteracted by repeating the measurements with different orientations of the antennas.
There is also an uncertainty regarding the elevation angle \(\theta\), as the height of the ionospheric layer where the signals have been reflected. For the Amateur radio services,  this may not be a significant drawback, as you measure the antenna performance in the directions where most WSPR receivers are located. This, in turn,  are often those locations with which one is planning to communicate. 

\section{Acknowledgement}
Thanks go to Björn Ekelund at Ericsson Research for valuable comments on the manuscript. The assistance in the field measurements by my student Sambhav Malhotra, is gratefully acknowledged.


% \end{IEEEbiography}

\end{document}